# Investigation of electrons interaction in a superconductor

Iogann Tolbatov

*Physics and Engineering Department, Kuban State University, Krasnodar, Russia*

(talbot1038@mail.ru)

Investigating the interaction of electrons in a superconductor by means of a method of solitary waves of Korteweg - de Vries, we refute the claim of absence of "Cooper pairs" in a superconductor. We also indicate that the nondissipative transfer of energy in the superconductor is possible only with the help of a pair of electrons.

The presence of "Cooper pairs" in a superconductor is mathematically and physically based model [1, 2], which does not explain, however, the mechanism of high temperature superconductivity (HTSC). There are many studies either based on "paired" electrons [3, 4], or on other models. For example, on the basis of the Hubbard model, assuming the Coulomb repulsion in a single center, it has been proposed to use the so-called models of resonant valence bonds [5]. Luttinger tried to explain the phenomenon of high-temperature superconductors assuming the separation of spin and charge [6]. Anderson developed the Luttinger fluid ideas as the essence of electronic systems in high-temperature superconductors [7, 8]. In the works of Laughlin [9], the fractional statistics was used for to describe the low-energy excitations in HTSC systems. Spin fluctuations [10, 11], the formation of "spin bag" [12, 13], the band structures possessing the "nesting" [14] were considered as the mechanisms leading to the electron-electron attraction and coupling.

Using the model of superconductivity [4], we investigate the superconductor (the general case that includes high- and low-temperature superconductivity cases) for the presence of "Cooper pairs".

As it follows from [4], the equations of motion of two quasi-free electrons forming a "Cooper pair" are functions of the coordinates of the two deformation potentials.

$$\frac{dx}{dt} = \frac{ieE_0}{m\omega_p} \sqrt{1-\varepsilon_1(x,y)} \exp\left(-\frac{i\omega_p t}{\sqrt{1-\varepsilon_1(x,y)}}\right),$$

$$\frac{dy}{dt} = \frac{ieE_0}{m\omega_p} \sqrt{1-\varepsilon_2(x,y)} \exp\left(-\frac{i\omega_p t}{\sqrt{1-\varepsilon_2(x,y)}}\right).$$

These deformation potentials determine the local values of the permittivities $\varepsilon_1(x,y)$ and $\varepsilon_2(x,y)$. Their electron localization coordinates are $x$ and $y$.

Moreover, $\frac{dx}{dt} = P(x,y)$; $\frac{dy}{dt} = Q(x,y)$; (1)

where $x = x(t)$, $y = y(t)$ — particle trajectories, $P(x,y)$, $Q(x,y)$ — analytical functions.

In the model [4], a pair of electrons is considered. We assume the independence of each electron, i. e. the absence of "Cooper pair". This corresponds to the condition:

$$\frac{dx}{dt} = P(x) \text{ and } \frac{dy}{dt} = Q(y). \qquad (2)$$

Due to the independence of each of the equations (2), it is more appropriate to consider the case:

$$\frac{dx}{dt} = P(x), \qquad (3)$$

$$\frac{dx}{dt} = \frac{ieE_0}{mw_p} \sqrt{1-\varepsilon(x)} \exp\left(-\frac{i\omega_p t}{\sqrt{1-\varepsilon(x)}}\right). \qquad (4)$$

Soliton is the solution of the nonlinear evolution equation, which at any given time is localized in some region of space, and the size of the region over time is limited, and the motion of the center region can be interpreted as the motion of a particle [15]. In the state of superconductivity, dissipations, as suggested in [4], do not exist or are replenished. Consequently, this process is possible for to describe the soliton. Densities of energy and momentum remain localized in the neighborhood of a point in space at any time. Localization takes place near the closed trajectory [4]. Such a localized solution is the kink-soliton [16].

We assume that the Korteweg - de Vries equation operating with only two independent variables $x$ and $t$ helps us to describe a superconducting system without the paired electrons - "Cooper pairs" (3).

Soliton of Korteweg - de Vries [17]

$$H_t + HH_x + H_{xxx} = 0 \tag{5}$$

describes a solitary wave

$$H_s(x,t) = \frac{3v}{ch^2\left[v^{1/2}\frac{1}{2}(x-vt)\right]}, \tag{6}$$

and it is uniquely determined by the following parameters:

a) velocity $v > 0$,

b) position of the maximum at a fixed time $t = 0$, $x = x_0$.

We denote: $v = \frac{dx}{dt}$. It follows from (4) that $v = P(x)$. Substituting $P(x)$ in (6) instead of $H_s(x,t)$, we obtain

$$P(x) = P(x(t)) = \frac{3v}{ch^2\left[v^{1/2}\frac{1}{2}(x-vt)\right]}, \text{ i. e. } v = \frac{3v}{ch^2\left[v^{1/2}\frac{1}{2}(x-vt)\right]}.$$

We transform the resulting equation:

$$3 = ch^2\left[v^{1/2}\frac{1}{2}(x-vt)\right], \text{ since } chw = \frac{e^w + e^{-w}}{2}, \text{ then}$$

$$10 = \exp\left[\left(\frac{dx}{dt}\right)^{1/2}\left(x-t\frac{dx}{dt}\right)\right] + \exp\left[-\left(\frac{dx}{dt}\right)^{1/2}\left(x-t\frac{dx}{dt}\right)\right].$$

Solving the equation $e^u + e^{-u} = 10$ obtained from the previous one by substituting $u = \left(\frac{dx}{dt}\right)^{1/2}\left(x-t\frac{dx}{dt}\right)$, we obtain:

$$u = \left(\frac{dx}{dt}\right)^{1/2}\left(x - t\frac{dx}{dt}\right) = \ln\frac{10 \pm \sqrt{96}}{2}.$$

We denote $\frac{dx}{dt} = \dot{x}$ for our convenience. Then $\dot{x}^{1/2}x - t\dot{x}^{3/2} = \ln\frac{10 \pm \sqrt{96}}{2}$,

$c = \ln\frac{10 \pm \sqrt{96}}{2}.$

Denoting $z = \dot{x}^{1/2}$ and $z^3 = \dot{x}^{3/2}$, we define the algebraic equation of third degree $ay^3 + bx^2 + cx + d = 0$ in the canonical form $z^3 + pz + q = 0$, where the coefficients are:

$$p = -\frac{b^2}{3a^2} - \frac{c}{a}, \quad q = \frac{2b^3}{27a^3} - \frac{bc}{3a^2} + \frac{d}{a}.$$

$$z^3 + \left(-\frac{x}{t}\right)z + \left(\frac{c}{t}\right) = 0 \tag{7}$$

The solution of this equation is carried out by radicals of the Cardano formula [18]:

$$z = \sqrt[3]{-\frac{q}{2} + \sqrt{\frac{q^2}{4} + \frac{p^3}{27}}} + \sqrt[3]{-\frac{q}{2} - \sqrt{\frac{q^2}{4} + \frac{p^3}{27}}}. \tag{8}$$

Applying this formula, it is necessary for each of the three values of the cube root $\alpha = \sqrt[3]{-\frac{q}{2} + \sqrt{\frac{q^2}{4} + \frac{p^3}{27}}} = \sqrt[3]{-\frac{c}{2t} + \sqrt{\frac{c^2}{4t^2} + \frac{x^3}{27t^3}}}$

to take the value of the root

$$\beta = \sqrt[3]{-\frac{q}{2} - \sqrt{\frac{q^2}{4} + \frac{p^3}{27}}} = \sqrt[3]{-\frac{c}{2t} - \sqrt{\frac{c^2}{4t^2} + \frac{x^3}{27t^3}}},$$

for which the condition

$\alpha\beta = -\frac{p}{3}$ is true, in our case this is $\alpha\beta = \frac{x}{3t}$.

In (8), numbers $p$ and $q$ are any complex numbers. However, it follows from (7) that numbers $p$ and $q$ are real. Consequently, the sign of the discriminant equation

$$D = -27q^2 - 4p^3 = 108\left(\frac{q^2}{4} + \frac{p^3}{27}\right) \text{ (in general case)}$$

$$D = -108\left[\left(\frac{c}{2t}\right)^2 - \left(\frac{x}{3t}\right)^3\right] \text{ (in our case)} \quad (9)$$

influences on the property of the roots of equation (7) of being real or imaginary [19].

1) If $D > 0$, the all three roots of equation (7) are real and distinct, but in (8), roots are expressed in terms of cubic radicals with imaginary radical expressions. This case is the irreducible one, since, despite the fact that in this case the coefficients and the roots are real, the roots can not be expressed in terms of coefficients with the radicals of the real numbers.

2) If $D < 0$, the all three roots are distinct, at that one root is valid, and two others are the conjugate imaginary ones.

3) If $D = 0$, the all roots are real, and that is what we need. Whether $p = q = 0$, we have a triple root, and in the case $p \neq 0$, $q \neq 0$, we have one double and one single roots.

We consider the third case in more detail:

a) $x \neq 0$, $-\frac{x}{t} \neq 0$, hence, $p \neq 0$,

b) $c \neq 0$, $\frac{c}{t} \neq 0$, i. e. $q \neq 0$.

Now we consider the discriminant (9). From $D = 0$, it follows that

$$\left(\frac{c}{2t}\right)^2 = \left(\frac{x}{3t}\right)^3,$$

hence, without solving the equation (7), we obtain the functions $x(t)$ and $\dot{x}(t)$:

$$x = 3\frac{c^{2/3}}{4^{1/3}}t^{1/3} \quad (10)$$

$$\dot{x} = \frac{c^{2/3}}{4^{1/3}}t^{-2/3}. \quad (11)$$

We recall that $c = \ln\frac{10 \pm \sqrt{96}}{2}$; $|c| = 2{,}29243$; $4^{1/3} = 1{,}58740$;

$K = \frac{c^{2/3}}{4^{1/3}} = 1{,}09524$ (accuracy in the calculations up to the fifth decimal place).

Dependencies (10) and (11) are converted to the form:

$$x = 3Kt^{1/3}, \tag{12}$$

$$\dot{x} = Kt^{-2/3}. \tag{13}$$

Taking into account also (3), we find from (12) and (13) that the time $t$ is a constant, as well as parameters $x$ and $\dot{x}$, and equation (2) in this case is also degenerate. Thus, equations (2) and (4) corresponding to the condition of "Cooper pair" absence are invalid as the condition itself.

Thus, we have: a) transformed model (1) [4] into (2) and (4) by means of assumption of the nonexistence of "Cooper pairs" in the superconductor; b) used the Korteweg - de Vries equation (5) as characterizing the superconducting (dissipationless) state process; c) showed the necessity of using the idea of paired electrons in the matter in superconducting state, by means of the model [4] example describing the mechanisms and the low- and high- temperature superconductivity.

So, in our study:

1) the existence of solitons in matter in the superconducting state is justified;

2) the absence of solitary waves described by the Korteweg - de Vries in matter in the superconducting state is shown;

3) the existence of "Cooper pairs" in a superconductor is confirmed.